\documentclass[11pt]{article}
\usepackage{fullpage,amssymb,amsmath,hyperref,url,graphicx}
\usepackage{xcolor,listings}
\usepackage{framed}

 \newenvironment{BoxedProposition}
   { \colorlet{shadecolor}{black!10}\begin{shaded}\begin{proposition}}
   {\end{proposition}\end{shaded}}

\graphicspath{{./fig/}{./}}

\lstdefinelanguage{Julia}%
  {morekeywords={abstract,break,case,catch,const,continue,do,else,elseif,%
      end,export,false,for,function,immutable,import,importall,if,in,%
      macro,module,otherwise,quote,return,switch,true,try,type,typealias,%
      using,while},%
   sensitive=true,%
   morecomment=[l]\#,%
   morecomment=[n]{\#=}{=\#},%
   morestring=[s]{"}{"},%
   morestring=[m]{'}{'},%
}[keywords,comments,strings]%

\definecolor{mydarkblue}{rgb}{0,0.08,0.45}

\usepackage{scalerel}
\usepackage{tikz}
\usetikzlibrary{svg.path}
\definecolor{orcidlogocol}{HTML}{A6CE39}
\tikzset{
  orcidlogo/.pic={
    \fill[orcidlogocol] svg{M256,128c0,70.7-57.3,128-128,128C57.3,256,0,198.7,0,128C0,57.3,57.3,0,128,0C198.7,0,256,57.3,256,128z};
    \fill[white] svg{M86.3,186.2H70.9V79.1h15.4v48.4V186.2z}
                 svg{M108.9,79.1h41.6c39.6,0,57,28.3,57,53.6c0,27.5-21.5,53.6-56.8,53.6h-41.8V79.1z M124.3,172.4h24.5c34.9,0,42.9-26.5,42.9-39.7c0-21.5-13.7-39.7-43.7-39.7h-23.7V172.4z}
                 svg{M88.7,56.8c0,5.5-4.5,10.1-10.1,10.1c-5.6,0-10.1-4.6-10.1-10.1c0-5.6,4.5-10.1,10.1-10.1C84.2,46.7,88.7,51.3,88.7,56.8z};
  }
}

\newcommand\orcidicon[1]{\href{https://orcid.org/#1}{\mbox{\scalerel*{
\begin{tikzpicture}[yscale=-1,transform shape]
\pic{orcidlogo};
\end{tikzpicture}
}{|}}}}

\newtheorem{remark}{Remark}
\newtheorem{proposition}{Proposition}
\newenvironment{proof}{\paragraph{Proof:}}{\hfill$\square$}

\def\ds{{\mathrm{d}s}}
\def\dxi{{\mathrm{d}\xi}}
\def\st{\ :\ }
\def\Bi{\mathrm{Bi}}
\def\diag{\mathrm{diag}}

\def\std{\mathrm{std}}

\def\calX{\mathcal{X}}

\def\bbR{\mathbb{R}}

\def\KL{\mathrm{KL}}

\def\calX{\mathcal{X}}

\def\calP{\mathcal{P}}

\def\KL{\mathrm{KL}}

\def\bbH{\mathbb{H}}
\def\st{\ :\ }
\def\bbZ{\mathbb{Z}}
\def\bbC{\mathbb{C}}
\def\Im{\mathrm{Im}}
\def\tr{\mathrm{tr}}
\def\inner#1#2{\left\langle #1,#2\right\rangle}
\def\Cov{\mathrm{Cov}}

\def\calH{\mathcal{H}}
\def\calP{\mathcal{P}}
\def\thetaZ{{\theta}}

\title{On the Kullback-Leibler divergence between discrete normal distributions}

\date{}

\author{Frank Nielsen\orcidicon{0000-0001-5728-0726}\\ Sony Computer Science Laboratories Inc.\\ Tokyo, Japan}

\begin{document}
\maketitle

\begin{abstract}
Discrete normal distributions are defined as the distributions with prescribed means and covariance matrices which maximize entropy on the integer lattice  support. 
The set of discrete normal distributions form an exponential family with cumulant function related to the Riemann  theta function. 
In this paper, we present several formula for  common statistical divergences between discrete normal distributions including the Kullback-Leibler divergence. 
In particular, we describe an efficient approximation technique for calculating  the Kullback-Leibler divergence between discrete normal distributions via the R\'enyi $\alpha$-divergences or the projective $\gamma$-divergences.
\end{abstract}


\tableofcontents

%%%
\section{Introduction}
%%%

\subsection{The continuous exponential family of normal distributions}

The $d$-variate normal distribution $N(\mu,\Sigma)$ 
 is characterized as the unique continuous distribution defined on the support $\calX=\bbR^d$ with prescribed mean $\mu$ and covariance matrix $\Sigma$ which maximizes Shannon's differential entropy~\cite{CT-1999}. 
Let $\calP_d$ denotes the open cone of positive-definite matrices and $\Lambda=\{(\mu,\Sigma)\ :\ \mu\in\bbR^d, \Sigma\in\calP_d\}$  the parameter space of the normal distributions. 
The probability density function (pdf) of a multivariate normal distribution $N(\mu,\Sigma)$ with parameterization $\lambda=(\mu,\Sigma)\in\Lambda$ is
$$
q_\lambda(x)=p_{\mu,\Sigma}(x)=\frac{1}{(2 \pi)^{\frac{d}{2}} \sqrt{|\Sigma|}} \exp \left(-\frac{1}{2}\left(x-\mu\right)^{\top} 
\Sigma^{-1}\left(x-\mu\right)\right),\quad \lambda\in\Lambda, x\in\bbR^d,
$$
where $|\Sigma|$ denotes the determinant of the covariance matrix. 

The set of normal distributions forms an exponential family~\cite{EF-2009,EF-2014} with pdfs~\cite{JS-2019} written canonically as
\begin{eqnarray}
q_\rho(x) &=& \frac{1}{Z_\bbR(\rho)} \exp\left(x^\top \rho_1+\tr\left(-\frac{1}{2}xx^\top \rho_2\right)\right),\nonumber\\
%&=&   \exp\left(\underbrace{x^\top \rho_1-\frac{1}{2}x^\top \rho_2 x}_{\inner{\rho}{t(x)}} -\underbrace{\log Z_\bbR(\rho)}_{F_\bbR(\rho)}\right),\label{eq:mvnef}
\end{eqnarray}
where $\rho=\left(\rho_1=\Sigma^{-1}\mu,\rho_2=\Sigma^{-1}\right)$ are the natural parameters corresponding to the sufficient statistics 
$t(x)=\left(x,-\frac{1}{2}xx^\top\right)$, and 
$Z_\bbR(\rho)$ is the partition function which normalizes the
 positive unnormalized density: 
\begin{eqnarray}
\tilde{q}_\rho(x)&=&\int_{\bbR^d} \exp\left(x^\top \rho_1-\frac{1}{2}x^\top \rho_2 x\right) \mathrm{d}x= (2\pi)^{\frac{d}{2}} |\rho_2^{-1}|^{\frac{1}{2}}\exp\left(\frac{1}{2}\rho_1^\top\rho_2^{-1}\rho_1\right).
\end{eqnarray}
Notice that we used the invariance of the matrix trace under cyclic permutations to get the last equality of Eq.~\ref{eq:mvnef}.
The cumulant function\footnote{Also called log-normalizer or log-partition function. 
The  naming ``cumulant function'' stems from the fact that the cumulant generating function $m_{X}(u)=E[\exp(u^\top t(x))]$ of the normal is $m_{X}(u)=F_\bbR(\rho+u)-F_\bbR(\rho)$ for $X\sim q_\rho$.} $F_\bbR(\rho)=\log Z_\bbR(\rho)$ of the multivariate normal distributions is
$$
F_\bbR(\rho)=\frac{1}{2}\left(\rho_1^\top\rho_2^{-1}\rho_1-\log |\rho_2|+d\log(2\pi)\right).
$$ 
Thus the pdf of a normal distribution writes canonically as the pdf of an exponential family:
\begin{eqnarray}
q_\rho(x)  &=&   \exp\left(\underbrace{x^\top \rho_1-\frac{1}{2}x^\top \rho_2 x}_{\inner{\rho}{t(x)}} -\underbrace{\log Z_\bbR(\rho)}_{F_\bbR(\rho)}\right),\label{eq:mvnef}\\
q_\lambda(x)  &=&  \exp\left(\inner{\rho(\lambda)}{t(x)}\right)-\log Z_\bbR(\rho(\lambda)),
\end{eqnarray}
where
$\inner{\rho}{\rho'}$ is the following compound vector-matrix inner product between $\rho=(a,B)$ and $\rho'=(a',B')$ with $a,a'\in\bbR^d$ and $B,B'\in\calP_d$:
$$
\inner{\xi}{\xi'}=a^\top a'+\tr(B'B).
$$

\subsection{The set of discrete normal distributions as a discrete exponential family}

Similarly, the $d$-variate discrete normal distribution\footnote{The term ``discrete normal distribution'' was first mentioned in~\cite{lisman1972note}, page 22 (1972).}~\cite{lisman1972note,Kemp-1997,DiscreteGaussianThetaSiegel-2019} $N_{\bbZ}(\mu,\Sigma)$ (or discrete Gaussian distribution~\cite{aggarwal2015solving,karmakar2018constant}) is defined as 
the unique discrete distribution (Theorem 2.5 of~\cite{DiscreteGaussianThetaSiegel-2019}) defined on the integer lattice support $\calX=\bbZ^d$ with prescribed mean $\mu$ and covariance matrix $\Sigma$ which maximizes Shannon's entropy. 
Therefore the set of discrete normal distributions is a discrete exponential family with probability mass function (pmf) which can be written canonically as
\begin{equation}\label{eq:dNpmf}
p_\xi(l) =\frac{1}{Z_{\bbZ}(\xi)} \exp\left(2\pi \left(-\frac{1}{2}l^\top \xi_2 l +l^\top \xi_1\right)\right),\quad l\in\bbZ^d.
\end{equation}
The sufficient statistic\footnote{The canonical decomposition of exponential families is not unique. 
We may choose $t_s(x)=st(x)$ and $\xi_s=\frac{1}{s}\xi$ for any non-zero scalar $s$:
The inner product remains invariant: $\inner{t(x)}{\xi}=\inner{t_s(x)}{\xi_s}$. Here, we choose $s=2\pi$ in order to reveal the Riemann theta function.} is $t(x)=\left(2\pi x,-\pi xx^\top\right)$ but the natural parameter $\xi=(\xi_1,\xi_2)$ cannot be written easily as a function of the $\lambda=(\mu,\Sigma)\in\Lambda$ parameters, where $\mu:=E_{p_\xi}[x]$ and 
$\Sigma=\Cov_{p_\xi}[x]=E_{p_\xi}[(x-\mu)(x-\mu)^\top]$. 
It can be shown that the normalizer is related to the Riemann theta function $\theta_R$  (Eq. 21.2.1 of~\cite{NIST-2010})  as follows:
$$
Z_{\bbZ}(\xi)=\theta_R(-i\xi_1,i\xi_2),
$$
where the complex-valued theta function is the holomorphic function defined by its Fourier series  as follows:
\begin{eqnarray*}
\theta_R &:& \bbC^d \times \calH_d \rightarrow \bbC\\ 
  && \theta_R(z,\Omega):=\sum_{l \in \bbZ^{d}} \exp\left(2\pi i\left(\frac{1}{2} l^\top \Omega l +l^\top z\right)\right),
\end{eqnarray*}
where  $\calH_d$ denotes the Siegel upper space\footnote{Siegel upper space generalizes the Poincar\'e hyperbolic upper space~\cite{nielsen2020siegel} $\bbH=\{z=x+iy\in\bbC \st y>0\}=\calH_1$.}~\cite{siegel2014symplectic} of symmetric complex matrices with positive-definite imaginary part:
$$
\calH_d=\left\{ R\in M(d,\bbC) \st R=R^\top, \Im(R)\in\calP_d\right\},
$$ 
with $M(d,\bbC)$ denoting the set of $d\times d$ matrices with complex entries.
A matrix $R\in\calH_d$ is called a Riemann matrix.
A Riemann matrix can be associated to a plane algebraic curve (loci of the zero of complex polynomial $P(x,y)$ with $x,y\in\bbC$) via a compact Riemann surface~\cite{RiemannMatrix-2011,swierczewski2016computing}.

\begin{remark}
Notice that the parameterization $\underline{\lambda}=(\underline{\mu},\underline{\Sigma})$ of continuous normal distribution applied to the discrete normal distribution for the pmf:
$$
p_{\underline{\lambda}}(l)\propto {\exp\left(-\frac{1}{2}(l-\underline{\mu})^\top \underline{\Sigma}^{-1}(l-\underline{\mu}) \right)},\quad l\in\bbZ^d
$$
yields in general $E_{p_{\underline{\lambda}}}[X]\not=\underline{\lambda}$ and $\Cov_{p_{\underline{\lambda}}}[X]\not=\underline{\Sigma}$.

Navarro and Ruiz~\cite{NavarroRuiz-2005} used the parameterization $(a,b)$ to  express the univariate pmf as
$$
p_{a,b}(x)=\frac{\exp\left(-\frac{(x-b)^2}{2a^2}\right)}{c(a,b)},
$$
where $c(a,b):=\sum_{x\in\bbZ} \exp\left(-\frac{(x-b)^2}{2a^2}\right)$.
This expression shows that discrete normal distributions are symmetric around the unique mode $b$: $p(b-x)=p(b+x)$.
Moreover, when $b$ is an integer, we have $E_{p_{a,b}}[x](a,b)=b$, and $\sigma^2(a,b)=\mathrm{Var}_{p_{a,b}}[x](a,b)=a^3\frac{c'(a)}{c(a)}$ where $c(a)=c(a,b)$ for integers $b$~\cite{NavarroRuiz-2005} .
\end{remark}

In the remainder, Let us denote the partition function of the discrete normal distributions $N_\bbZ(\xi)$ by
\begin{eqnarray*}
\thetaZ&:& \bbR^d\times\calP_d \rightarrow \bbR_+\\
\xi &\rightarrow& \thetaZ(\xi):=\theta_R(-i\xi_1,i\xi_2)=\sum_{l \in \bbZ^d} \exp\left(2\pi \left(\frac{1}{2} l^\top \xi_2 l +l^\top \xi_1\right)\right),
\end{eqnarray*}
with the corresponding cumulant function $F_\bbZ(\xi)=\log \thetaZ(\xi)$.
Both the continuous and discrete normal distributions are minimal regular exponential families with open natural parameter spaces   and   linearly independent sufficient statistic functions $t_i$'s.  
The orders of the $\bbR$-pmf discrete normal distributions  and the   $\bbC$-pmf discrete normal distributions are $\frac{d(d+3)}{2}$ and $d(d+3)$, respectively. 
By definition, the standard discrete normal distribution has zero mean and unit variance: Its corresponding natural parameters $\xi_\std$ 
can be approximated numerically as $\xi_\std\simeq (0,0.1591549\times I)$~\cite{DiscreteGaussianThetaSiegel-2019}, where $I$ denotes the identity matrix.
Observe that it is fairly different from the natural parameter $\rho_\std=(0,I)$ of the continuous normal distribution.

\begin{figure} 
\centering
\begin{tabular}{cc}
\includegraphics[width=0.4\columnwidth]{dnormal1D-0p0.3.pdf}&
\includegraphics[width=0.4\columnwidth]{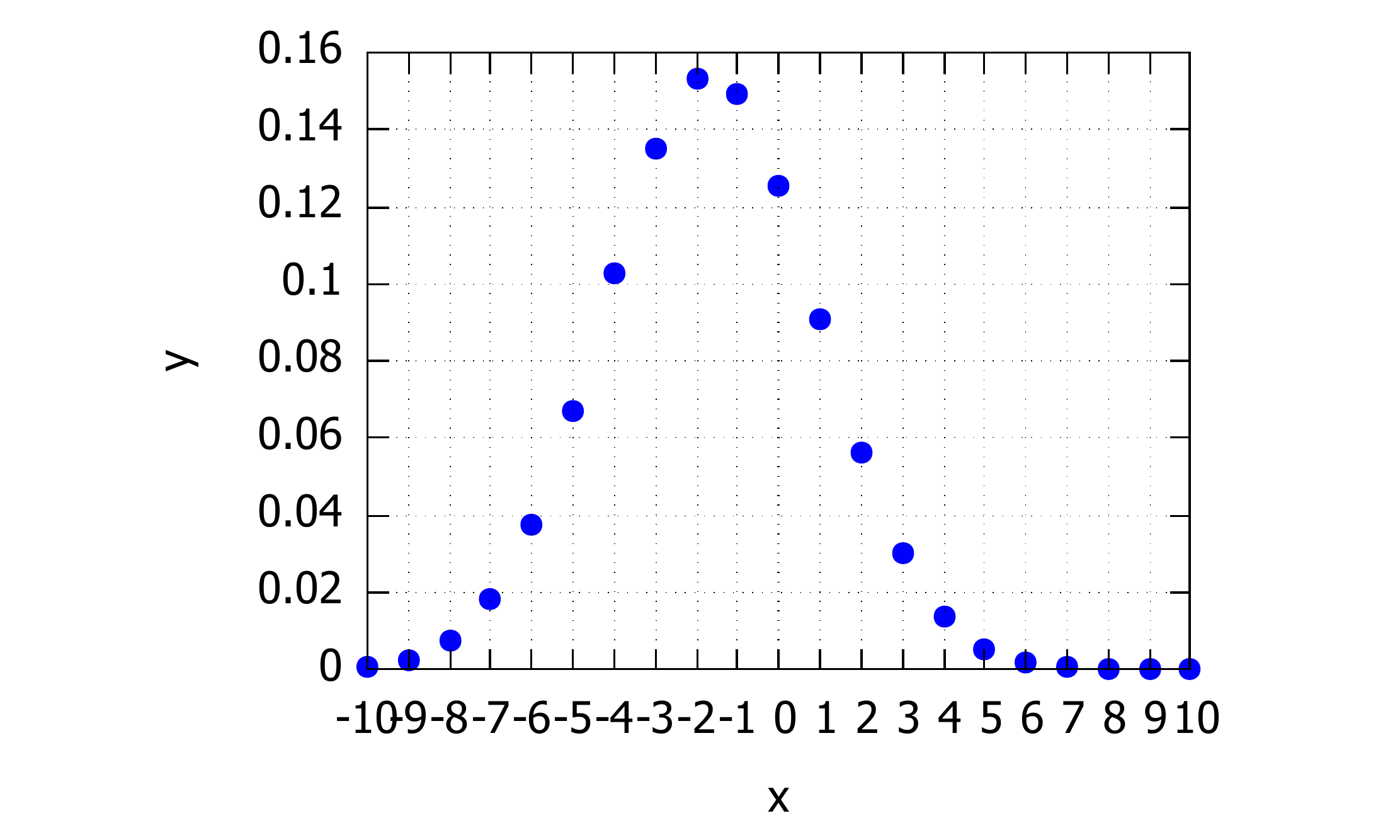}\\
\includegraphics[width=0.3\columnwidth]{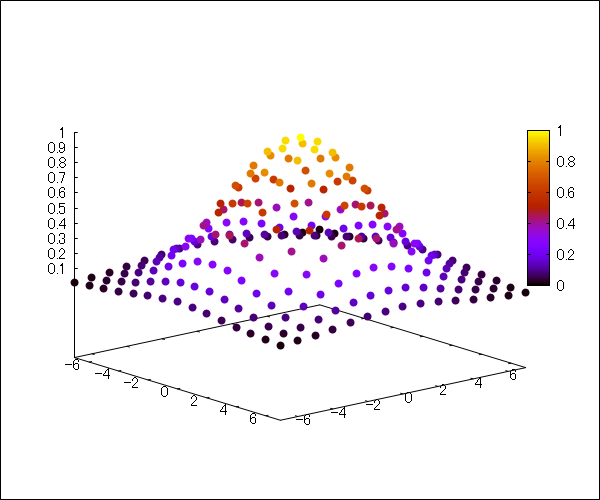}&
\includegraphics[width=0.3\columnwidth]{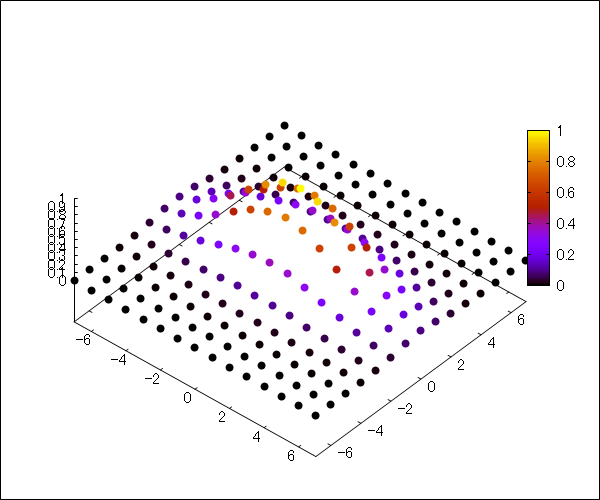}
\end{tabular}

\caption{Plot of unnormalized discrete normal distributions:
Top: $p_\xi$ on the 1D integer lattice $\bbZ$ clipped at $[-10,10]$ for $\xi=(0,0.3)$ (left) and $\xi=(0.25,0.15)$ (right).
Notice that when $\xi_1\in\bbZ$, the discrete normal is symmetric (left) but not for $\xi_1\not\in\bbZ$ (right).
Bottom: $\tilde{p}_\xi$ on the 2D integer lattice $\bbZ^2$ clipped at $[-7,7]\times[-7,7]$: 
(Left) $\xi_1=(0,0)$ and $\xi_2=\mathrm{diag}\left(\frac{1}{10},\frac{1}{10}\right)$,
(right) $\xi_1=(0,0)$ and $\xi_2=\mathrm{diag}\left(\frac{1}{10},\frac{1}{2}\right)$.}%
\label{fig:discreteN}%
\end{figure}

Let $p_\xi(x)=\frac{\tilde{p}_\xi(x)}{\thetaZ(\xi)}$,
with 
$$
\tilde{p}_\xi(x)=\exp\left(2\pi \left(-\frac{1}{2}x^\top \xi_2 x +x^\top \xi_1\right)\right).
$$
Figure~\ref{fig:discreteN} displays the plots of  two  unnormalized pmfs of two 1D discrete normal distributions and two 2D discrete normal distributions.

The discrete normal pmf of Eq.~\ref{eq:dNpmf} ($\bbR$-valued pmf) can be extended to complex-valued pmf\footnote{Complex-valued probabilities have been explored in quantum physics where the wave function can be interpreted as a complex-valued probability amplitude~\cite{youssef1994quantum}.} 
$p_{\zeta}^\bbC(l)$ ($\bbC$-pmf) when the parameter $\zeta$ belongs to the set
$\bbC^d\times \calH_d^\mathrm{right} \backslash \Theta_0$,
where $\calH_d^\mathrm{right}$ is Siegel right half-space (symmetric complex matrices with positive-definite real parts) and 
$$
\Theta_0=\{(a,B) \in \bbC^d\times \calH_d^\mathrm{right}  \st \theta_R(a,B)=0\},
$$ 
is called the universal theta divisor~\cite{de2010theta,DiscreteGaussianThetaSiegel-2019}. 
The zeros of the Riemann theta function\footnote{When $d=1$, the Riemann theta function is called the Jacobi
 theta function $\theta(z,\omega)=\sum_{l\in\bbZ} \exp(2\pi i lz+\pi il^2\omega)$. 
More precisely, we have $\theta_R(a,b)=\theta_3(\pi a,b)$ where $\theta_3$ denote the third-type of Jacobian theta function~\cite{NIST-2010}.} $\theta_R$ forms  an analytic variety of complex dimension $d-1$.
Notice both the probabilities and the parameter space of the complex discrete normal distribution are complex-valued ($\bbC$-pmf).
For example, consider $\zeta_1=(0,0)$ and $\zeta_2=(1+i)I$ (where $I$ denotes the identity matrix), then the $\bbC$-pmf evaluated at $l=(0,0)$ is 
$\frac{1}{\theta(\zeta)}\simeq\frac{1}{4+2i}$ which is a complex number.

The relationship between univariate discrete normal distributions and the Jacobi function was first reported in~\cite{Szablowski-2001}.
Studying more generally the $\bbC$-pmf discrete normal distributions using Siegel upper space $\calH_d^{\mathrm{right}}$ and Riemann theta function\footnote{By extending $\thetaZ$ to the Siegel right half-space.} allowed to get more easily results on the real-valued discrete normal distributions via properties of the theta function. For example, Agostini and Am\'endola~\cite{DiscreteGaussianThetaSiegel-2019} (Proposition 3.1) proved the quasiperiodicity\footnote{Namely the  Riemann theta function enjoys the following quasiperiodicity property:$\theta_R(z+u,\Omega)=\theta_R(z,\Omega)$ (periodic in $z$ with integer periods) and
 $\theta_R(z+\Omega v,\Omega)=\exp(-2\pi i(\frac{1}{2}v^\top \Omega v+v^\top z))\theta_R(z,\Omega)$ for any $u,v\in\bbZ^d$. The theta function can be generalized to the Riemann theta function with characteristic which involves a non-integer shift in its argument~\cite{swierczewski2016computing}.} of the complex discrete normal distributions
 $p_{a+iu+Bv,B}^\bbC(x) = p_{a,B}^\bbC(x-v)$ for any $(u,v)\in\bbZ^d\times\bbZ^d$.
We also have $p_{(a+\lambda B,B)}(x)=p_{(a,B)}{x-\lambda}$ and $p_{(a,B)}(x)=p_{(-a,B)(-x)}$ for (real) discrete normal distributions (parity property). 
Notice that the $\bbC$-pmf discrete normal distributions are not identifiable, i.e., $\zeta\mapsto p_{\zeta}$ is not one-to-one (Proposition 3.3~of~\cite{DiscreteGaussianThetaSiegel-2019}), but the $\bbR$-pmf discrete normal distributions are identifiable.

Table~\ref{tab:normal} displays the three types of normal distributions handled in this paper.

\begin{table}
\centering
{\footnotesize
\begin{tabular}{lllll}
Normal distribution &  support & natural parameter space & sufficient stats  & normalizer \\ \hline
$\bbR$-pdf continuous $q_\rho\sim N(\rho)$ & $\bbR^d$ & $\rho=\left(\Sigma^{-1}\mu,\Sigma^{-1}\right)\in\bbR^d\times\calP_d$ &  $\left(x,-\frac{1}{2}xx^\top\right)$ & $Z_\bbR(\rho)$\\
$\bbR$-pmf  discrete $p_\xi\sim N_\bbZ(\xi)$ & $\bbZ^d$ & $\xi\in\bbR^d\times \calP_d$ &   $\left(2\pi x,-\pi xx^\top\right)$& $\theta(\xi)=\theta_R(-i\xi_1,i\xi_2)$\\
$\bbR$-pmf  lattice  $p_\xi\sim N_{\Lambda}(\xi)$ & $\Lambda=L\bbZ^d$  & $\xi=(a,B)\in\bbR^d\times \calP_d$ &  $\left(2\pi x,-\pi xx^\top\right)$ & 
$\theta_\Lambda(\xi)=\theta_R(-iL^\top B L,iL^\top a)$\\
$\bbC$-pmf discrete   $p_i^\bbC\sim N_\bbZ(\zeta)$ & $\bbZ^d$ & $\zeta\in\bbC^d\times \calH^\mathrm{right}$ &  $\left(2\pi x,-\pi xx^\top\right)$ &$\theta_R(\zeta)$
\end{tabular}
}

\caption{Summary of the ordinary normal, discrete normal and $\bbC$-pmf discrete normal distributions viewed as natural exponential families.}\label{tab:normal}
\end{table}

A key property of Gaussian distributions is that the family is invariant under the action of affine automorphisms of $\bbR^d$.
Similarly, the family of discrete Gaussian distributions is invariant under the action of affine automorphisms of $\bbZ^d$
(Proposition 3.5~\cite{DiscreteGaussianThetaSiegel-2019}):
$$
	\forall \alpha\in\mathrm{GL}(d,\bbZ),\quad \alpha X_\xi=X_{\alpha^{-\top}\xi_1,\alpha^{-\top}\xi_2\alpha^{-1}}.
	$$
The parity property of discrete Gaussians follows  (Remark 3.7~\cite{DiscreteGaussianThetaSiegel-2019}):
$$
	X_{-\xi_1,\xi_2}\sim -X_\xi.
	$$

The discrete normal distributions play an important role as the counterpart of the normal distributions in robust implementations
on finite-precision arithmetic computers of algorithms in  differential privacy~\cite{wang2020d2p,DiscreteGaussianDiffPriv-2020} and lattice-based cryptography~\cite{budroni2021new}.
Recently, the discrete normal distributions have also been used in machine learning for a particular type of Boltzmann machine termed 
Riemann-Theta Boltzmann machines~\cite{carrazza2020sampling} (RTBMs). RTBMs have continuous visible states and discrete hidden states, 
 and the probability of hidden states follows a discrete multivariate Gaussian.

Let us mention that there exists other definitions of the discrete normal distributions.
For example, the discrete normal distribution may be obtained by quantizing the cumulative distribution function of the normal distribution~\cite{roy2003discrete}). This approach is also taken when considering mixtures of discrete normal distributions in~\cite{nichols2007automatic}.

%%%
\subsection{Discrete normal distributions on full-rank lattices}
%%%

\begin{figure}%
\centering
\includegraphics[width=0.4\columnwidth]{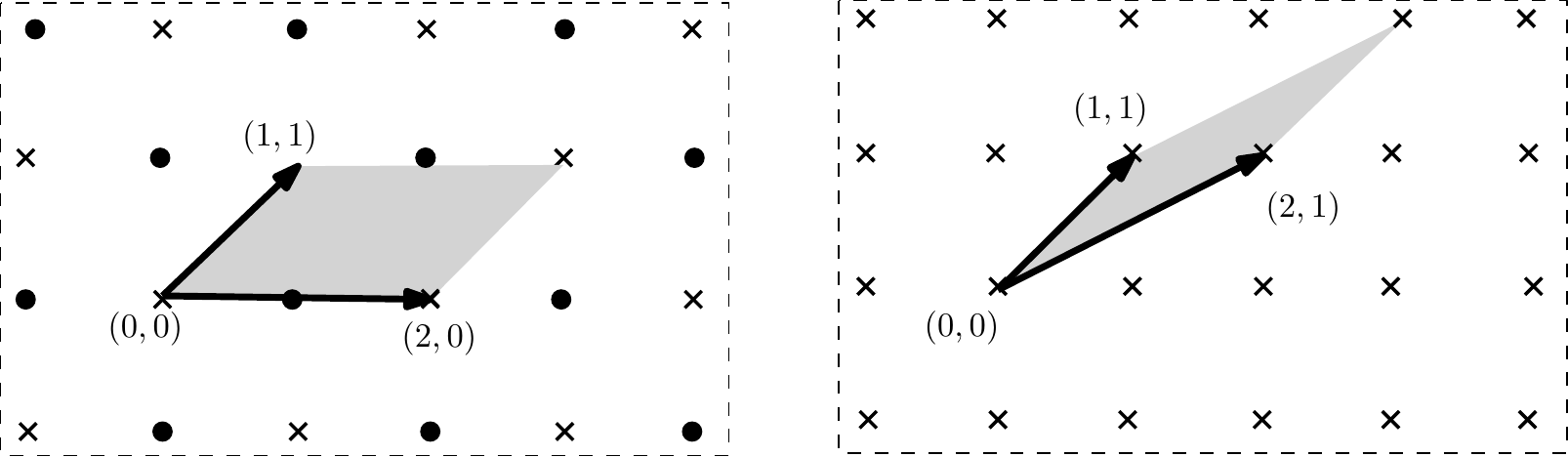}\vskip 0.3cm
\includegraphics[width=0.4\columnwidth]{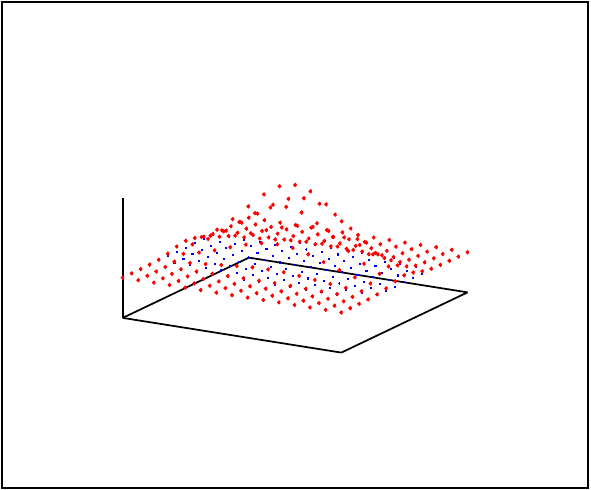}%
\caption{Top: Two examples of lattices with their basis defining a fundamental parallepiped: the left one yields a subset of $\bbZ^2$ while the second one coincides with $\bbZ^2$. 
Bottom: Lattice Gaussian $N_\Lambda(\xi)$ with $\Lambda=L\bbZ^2$ for $L=\left[\begin{array}{cc}1 & 1\cr 0 & 1\end{array}\right]$, and $\xi_1=(0,0)$ and $\xi_2=\diag(0.1,0.5)$. The lattice points are displayed in blue and the unnormalized pmf values at the lattice points are shown in red.}%
\label{fig:latticeGaussian}%
\end{figure}

Discrete normal distributions can also be defined on a $d$-dimensional lattice $\Lambda$ (also called full-rank lattice Gaussian distributions or lattice Gaussian measures with support not necessarily the integer lattice~\cite{gentry2008trapdoors,ling2014achieving} $\bbZ^d$) by choosing a set of linearly independent basis vectors $\{l_1,\ldots, l_d\}$ arranged in a basis matrix $L=[l_1,\ldots,l_d]$ and defining the lattice 
$\Lambda=L\bbZ^d=\{L\times l\ :\ l\in\bbZ^d\}$.
The pmf of a random variable $X\sim N_{\Lambda}(\xi)$ is
$$
p_{\xi}(x)=\frac{1}{\theta_\Lambda(\xi)}\exp\left(2\pi \left(-\frac{1}{2}x^\top \xi_2 x +x^\top \xi_1\right)\right),\quad x\in\Lambda.
$$
The above pmf can further be specialized for a random variable $X\sim N_{\Lambda}(c,\sigma)$ (a lattice Gaussian with variance $\sigma^2$ and center $c$) is 
$p_\xi(l)=\frac{1}{(\sqrt{2\pi}\sigma)^d}\exp(-\frac{\|l-c\|_2^2}{\sigma^2})$.
For a general lattice $\Lambda=L\bbZ^d$, we may define the lattice Gaussian distribution $N_\Lambda(\xi)$ with $\xi=(a,B)$ and normalizer
$$
\theta_\Lambda(\xi):=\sum_{l\in \Lambda} \exp\left(2\pi \left(-\frac{1}{2}l^\top \xi_2 l +l^\top \xi_1\right)\right).
$$ 
When $L=I$ (identity matrix), the lattice Gaussian distributions are the discrete normal distributions but other non-identity matrix basis may also generate $\bbZ^2$ (see Figure~\ref{fig:latticeGaussian}).
Since $\theta_\Lambda(\xi):=\sum_{l\in \bbZ^d} \exp\left(2\pi \left(-\frac{1}{2}(Ll)^\top \xi_2 (Ll) +(Ll)^\top \xi_1\right)\right)$,
 we have the following proposition:

\begin{BoxedProposition}
The normalizer of a lattice normal distribution $N_\Lambda(\xi)$ for $\Lambda=L\bbZ^d$ and $\xi=(a,B)$ amounts to the following Riemann theta function:
$$
\theta_\Lambda(\xi)=\theta_R(-iL^\top a,iL^\top B L).
$$
\end{BoxedProposition}

Last, we can translate the lattice $L\bbZ^d$ by $c\in\bbR^d$ (i.e., $\Lambda=L\bbZ^d+c$) so that we have the full generic pmf of a lattice gaussian which can be written for $\xi=(a,B)$ as:
\begin{eqnarray}
p_\xi^\Lambda(l) = \frac{1}{\theta_\Lambda(\xi)} \exp\left(2\pi \left(-\frac{1}{2}l^\top \xi_2 l +l^\top \xi_1\right)\right),\quad l\in L\bbZ^d+c,
\end{eqnarray}
where
\begin{eqnarray}
\theta_\Lambda(\xi) &=& \sum_{l\in\Lambda=L\bbZ^d+c}  \exp\left(2\pi \left(-\frac{1}{2}l^\top \xi_2 l +l^\top \xi_1\right)\right),\\
&=& \sum_{z\in\bbZ^d}  \exp\left(2\pi \left(-\frac{1}{2}(Lz+c)^\top \xi_2 (Lz+c) +(Lz+c)^\top \xi_1\right)\right).
\end{eqnarray}

The normalizer is related to the Riemann Theta functions with characteristics~\cite{NIST-2010} $\alpha,\beta\in\bbR^d$:

\begin{eqnarray*}
\theta_R\left[\begin{array}{c}
{\alpha} \\
{\beta}
\end{array}\right]({a}, {B}) &:=& \sum_{{l} \in \mathbb{Z}^{d}} e^{2 \pi i\left(\frac{1}{2}({l}+{\alpha})^\top {B}
 ({l}+{\alpha})+({l}+{\alpha})^\top ({B}+{\beta})\right)},\\
&=&
e^{2 \pi i\left(\frac{1}{2}  {\alpha}^T  {B} {\alpha}+ {\alpha}^\top ( {a}+ {\beta})\right)}\, \theta_R( {a}+ {B}  {\alpha}+ {\beta} ,  {B}).
\end{eqnarray*}

For example, when $L=I$, $\alpha=c$ and $\beta=0$.

\subsection{Contributions and paper outline}
%%%

We summarize our main contributions as follows: We report a formula for the R\'enyi $\alpha$-divergences between two discrete normal distributions in Proposition~\ref{prop:rd} including related results for the Bhattacharyya divergence, the Hellinger divergence and Amari's $\alpha$-divergences.
We give a formula for the cross-entropy between two discrete normal distributions in Proposition~\ref{prop:crossentropy} which yields a formula for the Kullback-Leibler divergence (Proposition~\ref{prop:klbd} and Proposition~\ref{prop:kldn}).
More generally, we extend the formula to Sharma-Mittal divergences in Proposition~\ref{prop:SM}.
In Section~\ref{sec:approx}, we show how to implement these formula using numerical approximations of the theta function.
We also propose a fast technique to approximate the Kullback-Leibler divergence between discrete normal distributions relying on $\gamma$-divergences~\cite{fujisawa2008robust} (Proposition~\ref{prop:approxKLgamma}).

\section{Statistical divergences between discrete normal distributions}
%%%%

%%%
\subsection{R\'enyi divergences}
%%%%

The R\'enyi $\alpha$-divergence~\cite{van2014renyi} between pmf $r(x)$ to pmf $s(x)$ on support $\calX$ is defined for any positive real $\alpha\not=1$ by
$$
D_{\alpha}[r:s] =\frac{1}{\alpha-1} \log \left(\sum_{x \in \calX} r(x)^{\alpha} s(x)^{1-\alpha}\right)= 
\frac{1}{\alpha-1} \log E_s\left[\left(\frac{r(s)}{s(x)}\right)^\alpha\right], \quad \alpha>0,\alpha\not=1.
$$

When $\alpha=\frac{1}{2}$, R\'enyi $\alpha$-divergence amounts to twice the symmetric  Bhattacharyya divergence~\cite{nielsen2011burbea}:
$D_{\frac{1}{2}}[r:s]=2\, D_{\mathrm{Bhattacharyya}}[r,s]$ with:
$$
D_{\mathrm{Bhattacharyya}}[r,s]:=-\log\left(\sum_{x \in \calX} \sqrt{r(x)s(x)} \right).
$$

The Bhattacharyya divergence can be interpreted as the negative logarithm of the  Bhattacharyya coefficient:
$$
\rho_{\mathrm{Bhattacharyya}}[r,s]=\sum_{x \in \calX} \sqrt{r(x)s(x)}.
$$
A divergence related to the Bhattacharyya divergence is the squared Hellinger divergence:
$$
D_{\mathrm{Hellinger}}^2[r,s]=\frac{1}{2}\sum_{x \in \calX} (\sqrt{r(x)}-\sqrt{s(x)})^2=1-\rho_{\mathrm{Bhattacharyya}}[r,s].
$$
The squared Hellinger divergence is one fourth of the $\alpha$-divergence for $\alpha=\frac{1}{2}$~\cite{IG-2016}, where
the $\alpha$-divergences are defined by
$$
D_{\mathrm{Amari},\alpha}[r:s]=\frac{1}{\alpha(1-\alpha)}\left(1-\rho_{\mathrm{Bhattacharyya},\alpha}[r:s]\right).
$$

The $\alpha$-divergences can be calculated from the skewed Bhattacharyya coefficients for $\alpha\in\bbR\backslash\{0,1\}$:
$$
\rho_{\mathrm{Bhattacharyya},\alpha}[r:s]=\sum_{x \in \calX} r(x)^\alpha s(x)^{1-\alpha}.
$$

Proposition~5 of~\cite{DiscreteGaussianDiffPriv-2020} upper bounds the R\'enyi $\alpha$-divergence between discrete normal distributions with same variance $\sigma^2$ as:
$$
D_{\alpha}\left[N_{\bbZ}\left(\mu_1, \sigma^{2}\right) : {N}_{\bbZ}\left(\mu_2, \sigma^{2}\right)\right] \leq 
\alpha\, \frac{(\mu_1-\mu_2)^{2}}{2 \sigma^{2}}.  
$$
 R\'enyi $\alpha$-divergences are non-decreasing with $\alpha$~\cite{van2014renyi}.

When both pmfs are from the same discrete exponential families with log-normalizer $F(\xi)=\log \theta(\xi)$, the R\'enyi $\alpha$-divergence~\cite{nielsen2011r} amounts to a $\alpha$-skewed Jensen divergence~\cite{nielsen2011burbea} between the corresponding natural parameters:
$$
D_{\alpha}[p_{\xi}:p_{\xi'}]=\frac{1}{1-\alpha} J_{F,\alpha}(\xi:\xi'),
$$
where
$$
J_{F,\alpha}(\xi:\xi'):=\alpha F(\xi)+(1-\alpha) F(\xi')-F(\alpha\xi+(1-\alpha)\xi').
$$

Indeed, let 
$$
I_{\alpha,\beta}[r:s]=
\sum_{x \in \calX} r(x)^\alpha s(x)^\beta ,
\quad \alpha,\beta\in\bbR.
$$
Then we have the following lemma:
\begin{BoxedProposition}\label{prop:Ialphabeta}
For two pmfs $p_\xi$ and $p_{\xi'}$ of a discrete exponential family with log-normalizer $F(\xi)$ with
$\alpha\xi+\beta\xi'\in\Xi$, we have
$$
I_{\alpha,\beta}[p_\xi:p_{\xi'}]=\exp\left(
F(\alpha\xi+\beta\xi')-(\alpha F(\xi)+\beta F(\xi'))
\right).
$$
\end{BoxedProposition}

\begin{proof}
We have
\begin{eqnarray*}
I_{\alpha,\beta}[p_\xi:p_{\xi'}]&=&
\sum_{x \in \calX} \exp(\inner{t(x)}{\alpha\xi}-\alpha F(\xi))
\,
\exp(\inner{t(x)}{\beta\xi'}-\beta F(\xi')),\\
&=&e^{F(\alpha\xi+\beta\xi')-(\alpha F(\xi)+\beta F(\xi'))}
\underbrace{ \sum_{x \in \calX} e^{\inner{t(x)}{\alpha\xi+\beta\xi'}-F(\alpha\xi+\beta\xi')}}_{=1},
\end{eqnarray*}
since $\sum_{x \in \calX} p_{\alpha\xi+\beta\xi'}(x)=1$ 
when $\alpha\xi+\beta\xi'\in\Xi$.
\end{proof}

Thus we get the following proposition:

\begin{BoxedProposition}\label{prop:rd}
The R\'enyi $\alpha$-divergence between two discrete normal distributions $p_\xi$ and $p_{\xi'}$ for $\alpha>0$ and $\alpha\not=1$ is
\begin{equation}\label{eq:RenyiZNor}
D_{\alpha}[p_{\xi}:p_{\xi'}] = \frac{1}{1-\alpha} \left(\alpha\log  \frac{\theta(\xi)}{\theta(\alpha\xi+(1-\alpha)\xi')}+(1-\alpha) \log
\frac{\theta(\xi')}{\theta(\alpha\xi+(1-\alpha)\xi')}\right).
\end{equation}
\end{BoxedProposition}

\begin{figure}%
\centering
\includegraphics[width=0.35\columnwidth]{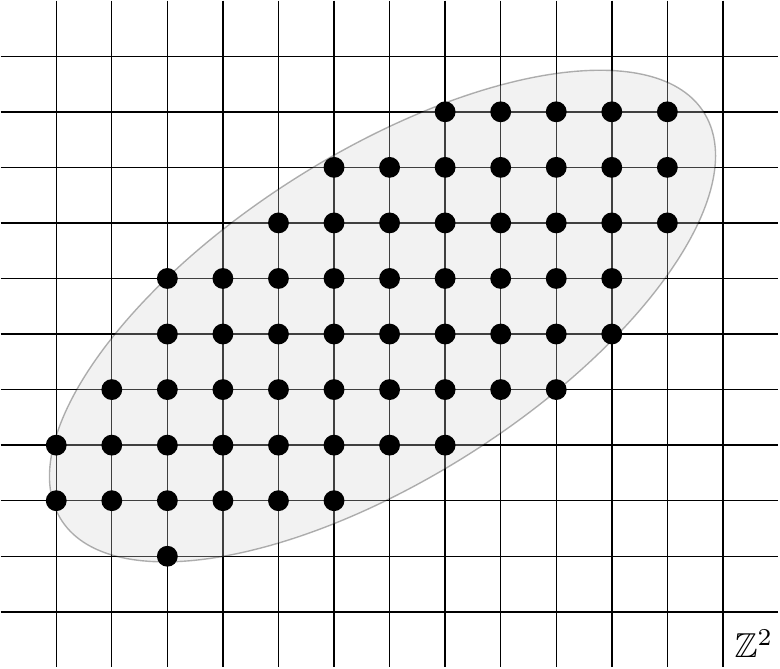}%
\caption{Approximating Riemann  $\theta_R$ function by summing on the integer lattice points falling inside an ellipsoid $E$: 
$\theta(\xi)\simeq\tilde{\theta}(\xi;E)$.}%
\label{fig:thetaellipsoid}%
\end{figure}

\begin{proof}

We have 
\begin{eqnarray*}
D_{\alpha}[p_{\xi}:p_{\xi'}]&=& \frac{1}{1-\alpha} \left(\alpha \log \theta(\xi)+(1-\alpha) \log\theta(\xi')-\log \theta(\alpha\xi+(1-\alpha)\xi')\right).
\end{eqnarray*}
Plugging $\log \theta(\alpha\xi+(1-\alpha)\xi')=(\alpha+1-\alpha)\log \theta(\alpha\xi+(1-\alpha)\xi')$ in the right-hand-side equation yields the result.
Notice that we can express also the R\'enyi divergences as
$$
D_{\alpha}[p_{\xi}:p_{\xi'}] =  \frac{1}{1-\alpha}\log \frac{\theta(\xi)^{\alpha}\theta(\xi')^{1-\alpha}}{\theta(\alpha\xi+(1-\alpha)\xi')}.
$$

See~\cite{deconinck2004computing,frauendiener2019efficient,agostini2021computing} for the  efficient  numerical  approximations of the Riemann theta function. Basically, the infinite theta series $\theta(\eta)$ is approximated by a finite summation over a region $R$ of integer lattice points:
$$
\tilde\theta(\xi;R):= \sum_{x\in R} \exp\left(2\pi \left(-\frac{1}{2}x^\top \xi_2 x +x^\top \xi_1\right)\right).
$$
When $R=\bbZ_d$, we have $\tilde\theta(\xi;R)=\theta(\xi)$.
The method proposed in~\cite{deconinck2004computing} consists in choosing the integer lattice points $E_\xi$  falling inside
 an ellipsoid used to approximate the theta function as illustrated in Figure~\ref{fig:thetaellipsoid}.
\end{proof}

Thus we have the following proposition:
\begin{BoxedProposition}
The squared Hellinger distance between two discrete normal distributions $p_{\xi}$ and $p_{\xi'}$ is
$$
D_{\mathrm{Hellinger}}^2[p_{\xi},p_{\xi'}]=1-\frac{\theta\left(\frac{\xi+\xi'}{2}\right)}{\sqrt{\theta(\xi)\theta(\xi')}}.
$$ 
\end{BoxedProposition}

We can also write
\begin{eqnarray*}
D_{\alpha}[p_{\xi}:p_{\xi'}]&=&\frac{1}{\alpha-1} \log E_{p_{\xi'}}\left[ \left( \frac{{p}_{\xi}}{{p}_{\xi'}} \right)^\alpha \right],\\
&=& \frac{1}{\alpha-1}  \left( \alpha\log\frac{\theta(\xi')}{\theta(\xi)} +E_{p_{\xi'}}\left[ \left( \frac{\tilde{p}_{\xi}(x)}{\tilde{p}_{\xi'}(x)} \right)^\alpha \right]  \right),\\
&=& \frac{\alpha}{\alpha-1}\log\frac{\theta(\xi')}{\theta(\xi)}  + \frac{1}{\theta(\xi')} \sum_{\in\bbZ^d} \tilde{p}_{\xi'}\left( \frac{\tilde{p}_{\xi}(x)}{\tilde{p}_{\xi'}(x)} \right)^\alpha.
\end{eqnarray*}
This last expression can be numerically estimated.

The Bhattacharyya  divergence between two discrete normal distributions $p_{\xi}$ and $p_{\xi'}$ can be expressed as an equivalent Jensen divergence between its natural parameters:
$$
D_{\mathrm{Bhattacharyya }}[p_{\xi'},p_{\xi'}]=J_F(\xi,\xi'),
$$
where
$$
J_{F}(\xi:\xi'):=\frac{F(\xi)+F(\xi')}{2}-F\left(\frac{\xi+\xi'}{2}\right).
$$
Thus we have $D_{\mathrm{Bhattacharyya }}[p_{\xi},p_{\xi'}]=\log\frac{\sqrt{\theta(\xi)\theta(\xi')}}{\theta\left(\frac{\theta(\xi)+\theta(\xi')}{2}\right)}$.
We can also express the  Bhattacharyya  divergence  using the unnormalized pmfs:
$$
D_{\mathrm{Bhattacharyya }}[p_{\xi'},p_{\xi'}]=\log\sqrt{\theta(\xi)\theta(\xi')}
-\log \left(\sum_{l\in\bbZ^d} \sqrt{\tilde{p}_{\xi}(l)\tilde{p}_{\xi'}(l)}\right).
$$

Consider the transformations $\tau$ that leaves the $\theta$ function invariant: $\theta(\tau(\xi))=\theta(\xi)$.
Then the R\'enyi $\alpha$-divergences simplifies to the following formula:

\begin{equation}\label{eq:rddninv}
D_{\alpha}[p_{\xi}:p_{\tau(\xi)}] = \frac{1}{1-\alpha}  \log \frac{\theta(\xi)}{\theta(\alpha\xi+(1-\alpha)\tau(\xi))}.
\end{equation}
For example, consider $\xi_1=\xi_1'\in\bbZ^d$ and $\xi_2=\diag(b_1,\ldots,b_d)$ and $\xi_2'=\diag(\sigma(b_1,\ldots,b_d))$ for a permutation $\sigma\in S_d$.  
Then we have $\theta(\xi')=\theta(\xi)$, and formula of Eq.~\ref{eq:rddninv} applies.

%%%
\subsection{Kullback-Leibler divergence: Dual natural and moment parameterizations}
%%%

When $\alpha\rightarrow 1$, the R\'enyi $\alpha$-divergences tend asymptotically to the Kullback-Leibler divergence (KLD).
The KLD between two pmfs $r(x)$ and $s(x)$ defined on the support $\calX$ is defined by
$$
D_{\KL}[r:s] = \sum_{x \in \calX} r(x)\log\frac{r(x)}{s(x)}.
$$

In general, the KLD between two pmfs of a discrete exponential family amounts to a reverse Bregman divergence between their natural parameters~\cite{nielsen2020elementary}:
$$
D_{\KL}[p_{\xi}:p_{\xi'}]={B_F}^*(\xi:\xi')=B_F(\xi':\xi),
$$
where the Bregman divergence with generator $F(\xi)$ is defined by:
$$
B_F(\xi':\xi) = F(\xi')-F(\xi)-\inner{\xi'-\xi}{\nabla F(\xi)},
$$
where $\inner{\xi}{\xi'}$ is the following compound vector-matrix inner product between $\xi=(a,B)$ and $\xi'=(a',B')$ with $a,a'\in\bbR^d$ and $B,B'\in\calP_d$:
$$
\inner{\xi}{\xi'}=a^\top a'+\tr(B'B).
$$

The gradient $\nabla F(\xi)=\frac{\nabla \theta(\xi)}{\theta(\xi)}$ defined the 
dual parameter $\eta$ of an exponential family: $\eta=\nabla F(\xi)$.
This dual parameter is called the moment parameter (or the expectation parameter)
because we have $E_{p_\xi}[t(x)]=\nabla F(\xi)$ and therefore $\eta=E_{p_\xi}[t(x)]$.
A discrete normal distribution can thus be parameterized either by its ordinary parameter $\lambda=(\mu,\Sigma)$,  its natural parameter $\xi$, or its  dual moment parameter $\eta$. We write the distributions accordingly: 
$N_\bbZ(\lambda)$, $N_\bbZ(\xi)$, and $N_\bbZ(\eta)$ with corresponding  pmfs: $p_\lambda(x)$, $p_\xi(x)$, and $p_\eta(x)$.

There exists a bijection between the space of natural parameters and the space of moment parameters induced by the Legendre-Fenchel transformation of the cumulant function:
$$
F^*(\eta)=\sup_{\xi\in\Xi} \inner{\xi}{\eta}-F(\theta),
$$
where $\Xi=\bbR^d\times\calP_d$.
Function $F^*$ is called the convex conjugate and induces a dual Bregman divergence so that we have $B_F(\xi':\xi)=B_{F^*}(\eta:\eta')$ with $\eta'=\nabla F(\xi')$.
The dual parameters are linked as follows:
$\eta=\nabla F(\xi)$, $\xi=\nabla F^*(\eta)$, and therefore we get:
$$
F^*(\eta)=\inner{\xi}{\eta}-F(\xi).
$$

The convex conjugate of the cumulant function $F(\xi)$ is called the negentropy because it can be shown~\cite{EF-2014,nielsen2010entropies} that we have
$$
F^*(\eta)=-H[p_\xi]=\sum_{x\in\calX} p_\xi(x)\log p_\xi(x),
$$
where $H[p_\xi]=-\sum_{x\in\calX} p_\xi(x)\log p_\xi(x)$ denotes Shannon's entropy of the random variable $X\sim p_\xi$.

The maximum likelihood estimator (MLE) of a density of an exponential family from $n$ identically and independently distributed samples $x_1,\ldots, x_n$ is given by~\cite{EF-2014}:
$$
\hat{\eta}=\frac{1}{n} \sum_{i=1}^n t(x_i).
$$
It follows from the equivariance property of the MLE that we have $\hat{\xi}=\nabla F^*(\hat{\eta})$.
We get the following MLE for the discrete normal family:

\begin{eqnarray*}
\hat{\eta}_1&=& \frac{2\pi}{n} \sum_{i=1}^n x_i = 2\pi\, \hat{\mu},\\
\hat{\eta}_2&=& -\pi \sum_{i=1}^n x_i x_i^\top= -\pi\, (\hat{\Sigma}+\hat{\mu}\hat{\mu}^\top).
\end{eqnarray*}

The Fenchel-Young inequality for convex conjugates $F(\xi)$ and $F^*(\eta)$ is
$$
F(\xi)+F^*(\eta')\geq \inner{\xi}{\eta'},
$$
with equality holding if and only if $\eta'=\nabla F(\xi)$.
The Fenchel-Young inequality induces a Fenchel-Young divergence:
$$
Y_{F,F^*}(\xi:\eta'):=F(\xi)+F^*(\eta')-\inner{\xi}{\eta'}=Y_{F^*,F}(\eta':\xi)\geq 0,
$$
such that $Y_{F,F^*}(\xi:\eta')=B_F(\xi:\xi')$.
Thus the  Kullback-Leibler divergence between two pmfs of a discrete exponential family can be expressed in the following equivalent ways using the natural/moment parameterizations:
\begin{equation}
D_{\KL}[p_{\xi}:p_{\xi'}]= B_F(\xi':\xi) = B_{F^*}(\eta:\eta')=Y_{F^*,F}(\eta:\xi') = Y_{F,F^*}(\xi':\eta).
\end{equation}

Thus using the fact that the KLD amounts to a reverse Bregman divergence for the cumulant function $F(\xi)=\log\theta(\xi)$, we get the following proposition:
 
\begin{BoxedProposition}\label{prop:klbd}
The Kullback-Leibler divergence between two discrete normal distributions $p_\xi$ and $p_\xi'$ with natural parameters $\xi$ and $\xi'$ is
$$
D_{\KL}[p_{\xi}:p_{\xi'}]=\log \frac{\theta(\xi')}{\theta(\xi)}-\frac{1}{\theta(\xi)}\inner{\xi'-\xi}{\nabla \theta(\xi)}.
$$
\end{BoxedProposition}

Some software packages for the Riemann theta function can numerically approximate both the theta function and its derivatives~\cite{agostini2021computing}.
Using the periodicity property of the theta function for $\xi'=(\xi_1+u,\xi_2)$ with $u\in\bbZ^d$, we have $\theta(\xi')=\theta(\xi)$, and therefore
$D_{\KL}[p_{\xi}:p_{\xi'}]=\frac{1}{\theta(\xi)}\inner{\xi-\xi'}{\nabla \theta(\xi)}$.

For the discrete normal distributions, we can express the moment parameter for the discrete normal distributions using the ordinary  mean-covariance parameters  $\lambda=(\mu,\Sigma)$.
Since the sufficient statistics is $2\pi(x,xx^\top)$, we have
$\eta_1(\xi)=E_{p_\xi}[2\pi x]=2\pi\mu$ and $\eta_2(\xi)=E_{p_\xi}[-\pi xx^\top]=-\pi(\Sigma+\mu\mu^\top)$.

Proposition~4.4 of~\cite{DiscreteGaussianThetaSiegel-2019}  reports the entropy of $p_\xi$ as
$$
H[p_\xi]=\log\theta(\xi)-2\pi \xi_1^\top\mu+\pi\,\tr(\xi_2(\Sigma+\mu\mu^\top)).
$$
We can rewrite the entropy as minus the convex conjugate function of the cumulant function:
$$
H[p_\xi]=-F^*(\eta)=F(\xi)-\inner{\xi}{\eta}.
$$
Thus we have the convex conjugate which can be expressed as
\begin{equation}
F^*(\eta)=-\log\theta(\xi)+2\pi \mu^\top \xi_1-\pi\, \tr(\xi_2(\Sigma+\mu\mu^\top)).  
\end{equation}

The entropy of $p_\xi$ can be calculated using the unnormalized pmf as follows:
$$
H[p_\xi]=-\sum_{l\in\bbZ^d} p_\xi(l)\log p_\xi(l)=-E_{p_\xi}[\log{p_\xi(l)}]
=\log\theta(\xi)-\frac{1}{\theta(\xi)} \sum_{l\in\bbZ^d} \tilde{p}_\xi(l)\log \tilde{p}_\xi(l) >0.
$$

The cross-entropy between two pmfs $r(x)$ and $s(x)$ defined over the support $\calX$ is
$$
H[r:s]=-\sum_{x\in\calX} r(x)\log s(x).
$$
Entropy is self cross-entropy: $H[r]=H[r:r]$.
The formula for the cross-entropy of a density of an exponential family~\cite{nielsen2010entropies} can be written as:
$$
H[p_\xi:p_{\xi'}]=F(\xi')-\inner{\xi'}{\nabla F(\xi)}=F(\xi')-\inner{\xi'}{\eta}.
$$

Thus we get the following proposition:

\begin{BoxedProposition}\label{prop:crossentropy}
The cross-entropy between two discrete normal distributions $p_\xi\sim N_\bbZ(\mu,\Sigma)$ and 
$p_{\xi'}\sim N_\bbZ(\mu',\Sigma')$ is
\begin{equation}
H[N_\bbZ(\mu,\Sigma):N_\bbZ(\mu',\Sigma')]=\log\theta(\xi')
-2\pi \mu^\top \xi_1'+\pi\, \tr(\xi_2'(\Sigma+\mu\mu^\top)).
\end{equation}
\end{BoxedProposition}

Notice that the cross-entropy can be written using the unnormalized pmf as
$$
H[p_{\xi}:p_{\xi'}]=-E_{p_{\xi}}[\log p_{\xi'}(x)] =\log\theta(\xi')
-\frac{1}{\theta(\xi)}\sum_{l\in\bbZ^d} \tilde{p}_\xi(l)\log \tilde{p}_{\xi'}(l).
$$

The KLD can be expressed as the cross-entropy minus the entropy (and henceforth its other name is relative entropy):
$$
D_\KL[p_{\xi}:p_{\xi'}]=H[p_{\xi}:p_{\xi'}]-H[p_{\xi}].
$$

It follows that we can compute the KLD between two discrete normal distributions as follows:

\begin{BoxedProposition}\label{prop:kldn}
The Kullback-Leibler divergence between two discrete normal distributions $p_\xi\sim N_\bbZ(\mu,\Sigma)$ and $p_\xi'\sim N_\bbZ(\mu',\Sigma')$ is:
\begin{equation}\label{eq:FYdrG}
D_\KL[p_\xi:p_{\xi'}]=\log \frac{\theta(\xi')}{\theta(\xi)} - 2\pi {\mu}^\top (\xi_1'-\xi_1) 
+\pi\, \tr((\xi_2'-\xi_2) (\Sigma+\mu\mu^\top)).
\end{equation}
\end{BoxedProposition}

Notice that we use mixed $(\xi,\lambda)$-parameterizations in the above formula. In practice, we estimate discrete normal distributions and then calculate the corresponding natural parameters by solving a gradient system explained in~\S\ref{sec:approx}.

Notice that the KLD between normal distributions can be decomposed as a sum of a squared Mahalanobis distance and a matrix Burg divergence (see Eq.~5 of~\cite{dhillon2007differential}).
For discrete normal distributions, when $\xi=(a,B)$ with $a\in\bbZ$ and $\xi'=(a+m,B)$ with $m\in\bbZ$, we have $\theta(\xi)=\theta(\xi')$ and $\mu=\mu'$, so that the KLD simplifies to the following formula:
$$
D_\KL[p_\xi:p_{\xi'}]=\frac{1}{\theta(\xi)}\inner{(-m,B)}{\nabla \theta(\xi)}.
$$

Notice that the MLE $\hat{\xi}_n$ of $n$  samples $x_1,\ldots,x_n\sim_{\mathrm{i.i.d.}} p_{\xi}$ can be interpreted as a KL divergence minimization problem:
$$
\hat{\xi}_n=\arg\min_{\xi\in\Xi} D_\KL[p_e:p_\xi],
$$
where $p_e(x)=\frac{1}{n} \sum_{i=1}^n \delta(x-x_i)$ denotes the empirical distribution with $\delta(x)$ the Dirac's distribution: 
$\delta(x)=1$ if and only if $x=0$.

Notice that when $\alpha\rightarrow 1$, we have $J_{F,\alpha}(\xi:\xi')\rightarrow B_F(\xi':\xi)$~\cite{nielsen2011burbea}, and 
$D_\alpha[p_\xi:p_{\xi}']\rightarrow D_\KL[p_\xi:p_{\xi}']]$.

\subsection{Sharma-Mittal divergences}
The Sharma-Mittal divergence~\cite{SM-2011} $D_{\alpha, \beta}[p:q]$ between two pmfs $p(x)$ and $q(x)$ defined over the discrete support $\calX$ unifies the R\'enyi $\alpha$-divergences ($\beta\rightarrow 1$) with the Tsallis $\alpha$-divergences ($\beta\rightarrow\alpha$):
$$
D_{\alpha, \beta}[p:q] := \frac{1}{\beta-1}\left(\left(\sum_{x\in\calX} p(x)^{\alpha} q(x)^{1-\alpha}\right)^{\frac{1-\beta}{1-\alpha}}-1\right),\quad \forall \alpha>0, \alpha \neq 1, \beta \neq 1.
$$
Moreover, we have $D_{\alpha, \beta}[p:q]\rightarrow D_\KL[p:q]$ when $\alpha,\beta\rightarrow 1$.

For two pmfs $p_{\xi}$ and $p_{\xi'}$ belonging to the same exponential family~\cite{SM-2011}, we have:
$$
D_{\alpha, \beta}[p_{\xi}:p_{\xi'}]=\frac{1}{\beta-1}\left(e^{-\frac{1-\beta}{1-\alpha} J_{F, \alpha}\left(\xi: \xi'\right)}-1\right).
$$

Thus we get the following proposition:

\begin{BoxedProposition}\label{prop:SM}
The Sharma-Mittal divergence $D_{\alpha, \beta}[p_{\xi}:p_{\xi'}]$ between two discrete normal distributions $p_\xi$ and $p_\xi'$ is:
\begin{eqnarray}
D_{\alpha, \beta}[p_{\xi}:p_{\xi'}] &=& \frac{1}{\beta-1}\left(\left(\frac{\theta_d(\xi)^{\alpha}\theta_d(\xi')^{1-\alpha}}{\theta_d(\alpha\xi+(1-\alpha)\xi')}\right)^{-\frac{1-\beta}{1-\alpha}}-1\right),\nonumber\\
&=& \frac{1}{\beta-1}\left( 
\left(\frac{\theta_d(\xi)}{\theta_d(\alpha\xi+(1-\alpha)\xi')}\right)^{\frac{\alpha(\beta-1)}{1-\alpha}}
  \left(\frac{\theta_d(\xi')}{\theta_d(\alpha\xi+(1-\alpha)\xi')}\right)^{\beta-1} 
\right).\label{eq:smdn}
\end{eqnarray}
\end{BoxedProposition}

%%%
\subsection{Chernoff information on the statistical manifold of discrete normal distributions}
%%%%

Chernoff information stems from the characterization of the error exponent in Bayesian hypothesis testing (see \S11.9 of~\cite{CT-1999}).
The Chernoff information between two pmfs $r(x)$ and $s(x)$ is defined by
$$
D_{\mathrm{Chernoff}}[r,s]:=-\min_{\alpha\in [0,1]} \log\left( \sum_{x\in\calX} r^{\alpha}(x) s^{1-\alpha}(x)\right),
$$
where $\alpha^*$ denotes the best exponent: $\alpha^*=\arg\min_{\alpha\in [0,1]} \sum_{x\in\calX} r^{\alpha}(x) s^{1-\alpha}(x)$.
When $r(x)=p_\xi(x)$ and $s(x)=p_{\xi'}(x)$ are pmfs of a discrete exponential family with cumulant function $F(\xi)$, we have  (Theorem~1 of~\cite{Chernoff-2013}):
$$
D_{\mathrm{Chernoff}}[p_\xi,p_{\xi'}]=B_F(\xi:\xi^*)=B_F(\xi':\xi^*),
$$
where $\xi^*:=\alpha^*\xi+(1-\alpha)\xi'$.
Thus calculating Chernoff information amounts to first find the best $\alpha^*$ and second compute $D_\KL[p_{\xi^*}:p_{\xi}]$ or equivalently $D_\KL[p_{\xi^*}:p_{\xi'}]$.
By modeling the exponential family as a manifold $M=\{p_\xi\ :\ \xi\in\Xi\}$ equipped with the Fisher information metric
 (a Hessian metric expressed in the $\xi$-coordinate system by $\nabla^2 F(\xi)$ 
so that the length element $\ds$ appears in the Taylor expansion of the KL divergence: 
$D_\KL[p_{\xi+\dxi}:p_{\xi}]=\frac{1}{2}\ds^2=\frac{1}{2}\dxi^\top \nabla^2 F(\xi)\dxi$), we can characterize geometrically the exact $\alpha^*$ (Theorem~2 of~\cite{Chernoff-2013}) as the unique intersection of an exponential geodesic $\gamma_{\xi,\xi'}$ with a mixture bisector $\Bi(\xi,\xi')$ where
\begin{eqnarray*}
\gamma_{\xi,\xi'} &:=& \{p_{\lambda\xi +(1-\lambda)\xi'}\propto p_\xi^\lambda p_{\xi'}^{1-\lambda}\ :\ \lambda\in (0,1)\},\\
\Bi(\xi,\xi')     &:=& \{p_{\omega}\in M\ :\ D_\KL[p_\omega:p_\xi]=D_\KL[p_\omega:p_{\xi'}]\}.
\end{eqnarray*}
Thus we have $p_{\xi^*}=\gamma_{\xi,\xi'}\cap\Bi(\xi,\xi')$.
This geometric characterization yields a fast numerical approximation bisection technique to obtain $\alpha^*$ within a prescribed precision error.
Since the discrete normal distributions form an exponential family, we can apply the above technique derived from information geometry\footnote{Information geometry is the field which considers differential-geometric structures of families of probability distributions.
 Historically, Hotelling~\cite{Hotelling-1930} first introduced the Fisher-Rao manifold. The term   ``information geometry'' occured in a paper of Chentsov~\cite{Chentsov-IG-1978} in 1978. } to calculate numerically the Chernoff information.
Various statistical inference procedures like estimators in curved exponential families and hypothesis testing can be investigated using the information-geometric dually flat structure of $M$, called a statistical manifold (see~\cite{IG-2016,nielsen2020elementary} for details).

\begin{remark}
The Fisher information matrix of the univariate normal distributions is $I(\xi)=(\log\theta(\xi))''=
\frac{\theta''(\xi)}{\theta(\xi)}-\left(\frac{\theta'(\xi)}{\theta(\xi)}\right)^2$, where $\theta'$ and $\theta''$ are the derivative and second derivatives of the Jacobi function $\theta$.
\end{remark}

Knowing that the KL divergence between two discrete normal distributions amounts to a Bregman divergence is helpful for a number of tasks like clustering~\cite{garcia2010simplification}:
The left-sided KL centroid of $n$ discrete normal distributions $p_{\xi_1},\ldots, p_{\xi_n}$ amounts to a right-sided Bregman centroid which is always the center of mass of the natural parameters~\cite{banerjee2005clustering}:
$$
\xi^*=\arg\min_\xi \sum_{i=1}^n \frac{1}{n} D_\KL[p_{\xi}:p_{\xi_i}]=\arg\min_\xi \sum_{i=1}^n \frac{1}{n} B_F({\xi_i}:\xi) \Rightarrow \xi^*=\frac{1}{n}\sum_{i=1}^n \xi_i.
$$

%%%
\section{Numerical approximations and estimations of divergences}\label{sec:approx}
%%%%

Although conceptually very similar as maximum entropy distributions to the continuous normal distributions, discrete normal distributions are mathematically very different to handle. On one hand, the normal distributions are exponential families with all parameter transformations and convex conjugates $F_\bbR(\rho)$ and $F_\bbR^*(\tau)$ available in closed-form~\cite{nielsen2021variational} (where $\tau=E_{q_{\rho}}[t(x)]$).
On the other hand, the discrete normal distributions with source parameters $\lambda=(\mu,\Sigma)$ can be converted from/back the moment parameters, but the conversions between natural parameters $\xi$ and expectation parameters $\eta=E_{p_\xi}[t(x)]=\nabla F(\xi)$ are not available in closed-form, nor the  cumulant function $F(\xi)=\log\theta(\xi)$ and its convex conjugate $F^(\eta)$.

%%%
\subsection{Converting numerically natural to moment parameters and vice versa}\label{sec:convertnaturalmoment} 
%%%%

In practice, we can approximate the  conversion procedures $\xi\leftrightarrow\eta$ as follows:
\begin{itemize}
	\item Given natural parameter $\xi$, we may approximate the dual moment parameter $\eta=\nabla F(\xi)=E_{p_xi}[t(x)]$ as $\tilde{\eta}=\frac{1}{m}\sum_{i=1}^n t(x_i)$ where $x_1,\ldots, x_m$ are independently and identically sampled from $N_{\bbZ^d}(\xi)$.
	Sampling uniformly from discrete normal distributions can be done exactly in 1D~\cite{DiscreteGaussianDiffPriv-2020} (requiring average constant time) but requires sampling heuristics in dimension $d>1$. 
	Two common sampling heuristics approximating for handling discrete normal distributions are:
	\begin{itemize}
		\item $H_1$: Draw a variate $x\sim q_{\mu,\Sigma}$ from the corresponding 
  normal distribution $q_{\mu,\Sigma}$, and round or choose the closest integer lattice point $\tilde{x}$ of $\bbZ^d$ with respect to the $\ell_1$-norm (i.e., $\tilde{x}=\arg\min_{l\in\bbZ^d} \|l-x\|_1=\sum_{i=1}^d |l^i-x^i|$), where $(l^1,\ldots, l^d)$ and $(x^1,\ldots,x^d)$ denote the coordinates of $l$ and $x$, respectively.

\item $H_2$: Consider the integer lattice points $E_\xi$  falling inside the ellipsoid region~\cite{deconinck2004computing} used for approximating $\theta(\xi)$ by $\tilde{\theta}(\xi;E_\xi)$ (Figure~\ref{fig:thetaellipsoid}), draw uniformly an integer lattice point $l$ from $E_\xi$ and accept it with probability $p_\xi(l)$ (acceptance-rejection sampling described in~\cite{carrazza2020sampling}).

	\end{itemize}

	\item Given the moment parameter $\eta$, we may approximate $\theta=\nabla F^*(\eta)$ by solving a gradient system.
	Since the moment generating function (MGF) of an exponential family~\cite{EF-2014} is 
$m_X(u):=E_X[\exp(u^\top X)]=\exp(F(\xi+u)-F(\xi))$,
we deduce that the MGF of the discrete normal distributions $X\sim p_\xi$ is
$$
m_\xi(u)=\frac{\theta(\xi+u)}{\theta(\xi)}.
$$
The non-central moments of the sufficient statistics (also called raw moments or geometric moments) of an exponential family can be retrieved from the partial derivatives of the MGF.
For the discrete normal distributions, Agostini and Am\'endola~\cite{DiscreteGaussianThetaSiegel-2019} obtained the following gradient system:
\begin{eqnarray*}
\eta_1 &=& E_{p_\xi}[t_1(x)]=\frac{1}{2\pi} \frac{1}{\theta(\xi)} \nabla_{\xi_1} \theta(\xi),\\
\eta_2 &=& E_{p_\xi}[t_2(x)]=-\frac{1}{2\pi} \frac{1}{\theta(\xi)} 
\left(\nabla_{\xi_2} \theta(\xi)+\mathrm{diag}(\nabla_{\xi_2} \theta_d(\xi))\right).
\end{eqnarray*}
In practice, this gradient system can be solved up to arbitrary machine precision using software packages (initialization can be done from the closed-form conversion of the moment parameter to the natural parameter for the continuous normal distribution).
For example, one way to solve the gradient system is by using the technique described in~\cite{zellner1988calculation} that we summarize as follows:

First, let us choose the following canonical parameterization of the densities of an exponential family:
$$
p_\psi(x):=\exp\left(-\sum_{i=0}^D \psi_i t_i(x)\right).
$$
That is, $\psi_0=F(\psi)=$ and $\psi_i=-\xi_i$ for $i\in\{1,\ldots, D\}$ (i.e., parameter $\psi$ is an augmented natural parameter which includes the log-normalizer in its first coefficient).

Let $K_i(\psi):=E_{p_\theta}[t_i(x)]=\eta_i$ denote the set of $D+1$ non-linear equations for $i\in\{0,\ldots, D\}$. 
The method of~\cite{zellner1988calculation}  converts iteratively $p^\eta$ to $p_\psi$. 
We initialize $\psi^{(0)}$ and calculate numerically $\psi_0^{(0)}=F(\psi^{(0)})$.

At iteration $t$ with current estimate $\psi^{(t)}$, we use the following first-order Taylor approximation:
$$
K_i(\psi)\approx K_i(\psi^{(t)})+(\psi-\psi^{(t)})\nabla K_i(\psi^{(t)}).
$$
Let $H(\psi)$ denote the $(D+1)\times (D+1)$ matrix:
$$
H(\psi):=\left[\frac{\partial K_i(\psi)}{\partial\psi_j}\right]_{ij}.
$$
We have
\begin{equation}\label{eq:Hmatrix}
H_{ij}(\psi)=H_{ji}(\psi)=-E_{p_\psi}[t_i(x)t_j(x)].
\end{equation}

We update as follows:
\begin{equation}\label{eq:update}
\psi^{(t+1)}=\psi^{(t)}+H^{-1}(\psi^{(t)})
\left[\begin{array}{c}
\eta_0-K_0(\psi^{(t)})\\ \vdots\\ \eta_D-K_D(\psi^{(t)}) \end{array}\right].
\end{equation}

When implementing this method, we need to approximate $H_{ij}$ of Eq.~\ref{eq:Hmatrix} using the theta ellipsoid points.
For $d$-variate discrete normal distributions with $D=\frac{d(d+3)}{2}$, we have $t_1(x)=x_1,\ldots, t_d(x)=x_d, 
t_{d+1}(x)=-\frac{1}{2}x_1x_1, t_{d+2}(x)=-\frac{1}{2}x_1x_2,\ldots, t_{D}(x)=-\frac{1}{2}x_dx_d$.

\end{itemize}

%%%
\subsection{Some illustrating numerical examples}\label{sec:numexample}
%%%%
To compute numerically the theta functions and its derivatives, we may use the following software packages (available in various  programming languages):
 {\tt abelfunctions} in {\tt SAGE}~\cite{abelfunctions},
{\tt algcurves} in {\tt Maple}\textregistered{}~\cite{algcurves-2011},
{\tt Theta} in {\tt Python}~\cite{carrazza10theta},
{\tt Riemann} of {\tt jTEM} (Java Tools for Experimental Mathematics) in {\tt Java}~\cite{hoffmann2006jreality} (see also~\cite{deconinck2004computing}), or
{\tt Theta.jl} in {\tt Julia}~\cite{agostini2021computing}.

\begin{figure}%
\centering
\includegraphics[width=0.45\columnwidth]{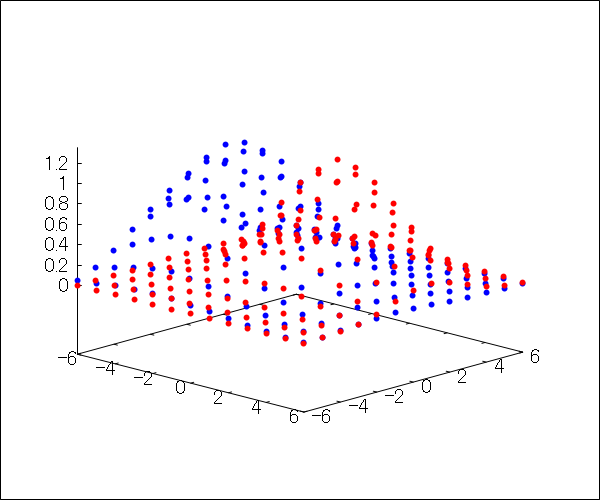}%
\caption{Two bivariate discrete normal distributions used to calculate statistical divergences.}%
\label{fig:dNex12}%
\end{figure}

For our experiments, we used {\tt Java}\texttrademark (in-house implementation) and {\tt Julia} (with the package {\tt Theta.jl}~\cite{agostini2021computing}).
We consider the following two discrete normal distributions $p_\xi$ and $p_{\xi'}$ with the following parameters:  
\begin{eqnarray*}
\xi&=&\left((-0.2,-0.2),\diag(0.1,0.2)\right),\\
\xi'&=&\left((0.2,0.2),\diag(0.15,0.25)\right).
\end{eqnarray*}
These bivariate discrete normal distributions are plotted in Figure~\ref{fig:dNex12}.

We implemented the statistical divergences between discrete normal distributions using an in-house Java\texttrademark{} software and Julia {\tt Theta.jl}~\cite{agostini2021computing} package (see Appendix~\ref{sec:Julia} for a code snippet).

For the above discrete normal distributions, we calculated:
$$
D_{\mathrm{Bhattacharyya}[p_\xi},p_{\xi'}]=\frac{1}{2}\,D_{\frac{1}{2}_\KL}[p_\xi:p_{\xi'}] \simeq 1.626,
$$
and approximated the KL divergence by the R\'enyi divergence for $\alpha_\KL=1-10^{-5}=0.99999$:
$$
D_{\KL}[p_\xi:p_{\xi'}]\simeq D_{\alpha_\KL}[p_\xi:p_{\xi'}] =  \frac{1}{1-\alpha_\KL} J_{F,\alpha_\KL}(\xi:\xi') \simeq 7.84.
$$

Implementing these formula required to calculate $F(\xi)$, i.e., to evaluate the logarithm of theta functions.
The following section describes another efficient method based on a projective divergence, i.e., a divergence which does not require pmfs to be normalized.

 %%%%%
\subsection{Approximating the Kullback-Leibler divergence via projective $\gamma$-divergences}\label{sec:gamma}
%%%%

The $\gamma$-divergences~\cite{fujisawa2008robust,cichocki2010families} between two pmfs $p(x)$ and $q(x)$ defined over the support $\calX$  for a real $\gamma>1$  is defined by:
$$
D_\gamma[p:q]
:=\frac{1}{\gamma(\gamma-1)} 
\log \left(
\frac{\left(\sum_{x\in\calX} p^{\gamma}(x)\right)\, \left(\sum_{x\in\calX} q^{\gamma}(x)\right)^{\gamma-1}}{\left(\sum_{x\in\calX} p(x) q^{\gamma-1}(x))
\right)^{\gamma}}\right),\quad (\gamma>1).
$$
The $\gamma$-divergences are projective divergences, i.e., they satisfy the following identity:  
$$
D_{\gamma}[p:p']=D_{\gamma}[\lambda p:\lambda'p'],\quad (\forall\lambda,\lambda'>0).
$$

Thus let us rewrite $p(x)=\frac{\tilde{p}(x)}{Z_p}$ and $q(x)=\frac{\tilde{q}(x)}{Z_q}$ where $\tilde{p}(x)$ and $\tilde{q}(x)$ are computationally tractable unnormalized pmfs, and $Z_p$ and $Z_q$ their respective computationally intractable normalizers.
Then we have 
$$
D_{\gamma}[p:p']=D_{\gamma}[\tilde{p}:\tilde{p}'].
$$

Let us define
$$
I_{\gamma}[{p}:{q}]:=\sum_{x \in \calX} {p}(x) {q}(x)^{\gamma-1}.
$$
Then the $\gamma$-divergence can be written  as:
$$
D_\gamma[p:q]=D_{\gamma}[\tilde{p}:\tilde{q}]
=\frac{1}{\gamma(\gamma-1)} 
\log \left(
\frac{I_\gamma[\tilde{p}:\tilde{p}]\, I_\gamma[\tilde{q}:\tilde{q}]^{\gamma-1}}{I_\gamma[\tilde{p}:\tilde{q}]^{\gamma}}\right).
$$

Consider $p=p_\xi$ and $q=p_{\xi'}$ two pmfs belonging to the lattice Gaussian exponential family, and let $$
\tilde{I}_{\gamma}\left(\xi:\xi'\right)=I_{\gamma}\left[\tilde{p}_{\xi}:\tilde{p}_{\xi'}\right].
$$

Provided that $\xi+(\gamma-1)\xi'\in\Xi$, we have following the proof of Proposition~\ref{prop:Ialphabeta} that
\begin{eqnarray*}
\tilde{I}_{\gamma}\left(\xi:\xi'\right)&=&
\sum_{l\in\Lambda} \tilde{p}_\xi(l)\tilde{p}_{\xi'}(l)^{\gamma-1},\\
&=&  \sum_{l\in\Lambda} \exp(\inner{\xi+(\gamma-1)\xi'}{t(x)}),\\
&=&\exp(F_\Lambda(\xi+(\gamma-1)\xi')) \, \underbrace{\sum_{l\in\Lambda} p_{\xi+(\gamma-1)\xi'}(l)}_{=1},\\
&=& \exp(F_\Lambda(\xi+(\gamma-1)\xi')),
\end{eqnarray*}
where $F_\Lambda(\xi)=\log\theta_\Lambda(\xi)$ denotes the cumulant function of the Gaussian distributions on lattice $\Lambda$.
That is, we have
$$
\tilde{I}_{\gamma}\left(\xi:\xi'\right)=\theta_\Lambda(\xi+(\gamma-1)\xi'),
$$
and therefore, we can express the $\gamma$-divergences as
\begin{equation}
D_\gamma[p_\xi:p_{\xi'}]
=
\frac{1}{\gamma(\gamma-1)} 
\log \left(
\frac{\theta_\Lambda(\gamma\xi)\, \theta_\Lambda(\gamma\xi')^{\gamma-1}}{
\theta_\Lambda(\xi+(\gamma-1)\xi')^{\gamma}
}\right).
\end{equation}

Notice that the exact values of the infinite summations $\tilde{I}_{\gamma}\left(\xi:\xi'\right)$  depend on the Riemannian theta function.

Now, the $\gamma$-divergences tend asymptotically to the Kullback-Leibler divergence between normalized densities when $\gamma\rightarrow 1$~\cite{fujisawa2008robust,cichocki2010families}:
 $\lim_{\gamma\rightarrow 1} D_\gamma[\tilde{p}:\tilde{q}]=D_\KL\left[\frac{\tilde{p}}{Z_p}:\frac{\tilde{q}}{Z_q}\right]$.
Let us notice that the KLD is not a projective divergence, and that for small enough $\gamma>1$, we have $\xi+(\gamma-1)\xi'$ always falling inside the natural parameter space $\Xi$.
Moreover, we can approximate the infinite summation  using a finite region of integer lattice points $R_{\xi,\xi'}$:
$$
\tilde{I}_{\gamma,R_{\xi,\xi'}}(\xi:{\xi'}):=\sum_{x \in R_{\xi,\xi'}} \tilde{p}_\xi\, \tilde{p}_{\xi'}(x)^{\gamma}.
$$

For example, we can use the theta ellipsoids~\cite{deconinck2004computing} $E_\xi$ and $E_{\xi'}$ used to approximate $\theta(\xi)$ and $\theta(\xi')$, respectively (Figure~\ref{fig:thetaellipsoid}): We choose $R_{\xi,\xi'}=(E_\xi\cup E_{\xi'})\cap\bbZ^d$.
In practice, this approximation of the $I_\gamma$ summations scales well in high dimensions.
Overall, we get our approximation of the KLD between two lattice Gaussian distributions summarized in the following proposition:

\begin{proposition}\label{prop:approxKLgamma} 
The Kullback-Leibler divergence between two lattice Gaussian distributions $p_\xi$ and $p_{\xi'}$ can be efficiently approximated:
\begin{equation}
D_\KL[p_\xi:p_{\xi'}]\approx D_\gamma[p_\xi:p_{\xi'}]
=
\frac{1}{\gamma(\gamma-1)} 
\log \left(
\frac{(\tilde{I}_{\gamma,R_\xi}(\xi:\xi)\, \tilde{I}_{\gamma,R_\xi'}(\xi':\xi')^{\gamma-1}}{\tilde{I}_{\gamma, R_{\xi,\xi'}}(\xi:\xi')^{\gamma}}\right),
\end{equation}
for $\gamma>1$ close to $1$ (say, $\gamma=1+10^{-5}$), where $R_\xi$ and $R_{\xi'}$ denote the integer lattice points falling inside the theta ellipsoids $E_\xi$ and $E_{\xi'}$ used to approximate the theta functions~\cite{deconinck2004computing} $\theta_\Lambda(\xi)$ and $\theta_\Lambda(\xi')$, respectively.
\end{proposition}

Table~\ref{tab:summarydiv} summarizes the various closed-formula obtained for the  statistical divergences between lattice Gaussian distributions considered in this paper.
 
{\footnotesize
\begin{table}
\centering

{%\footnotesize
\begin{tabular}{ll}
Divergence & definition/closed-form formula for lattice Gaussians\\ \hline
Kullback-Leibler divergence & $D_\KL[p_\xi:p_{\xi'}]=\sum_{l\in\Lambda} p_\xi(l)\log \frac{p_\xi(l)}{p_{\xi'}(l)}$ \\
&
$D_\KL[p_\xi:p_{\xi'}]=\log\left( \frac{\theta_\Lambda(\xi')}{\theta_\Lambda(\xi)}\right)$\\
& \hskip 2cm $- 2\pi {\mu}^\top (\xi_1'-\xi_1) 
+\pi\, \tr\left((\xi_2'-\xi_2) (\Sigma+\mu\mu^\top)\right)$\\
squared Hellinger divergence  & $D_{\mathrm{Hellinger}}^2[p_\xi:p_{\xi'}]=\frac{1}{2}\sum_{l \in \Lambda} (\sqrt{p_\xi(l)}-\sqrt{p_{\xi'}(l)})^2$ \\ &
$D_{\mathrm{Hellinger}}^2[p_\xi:p_{\xi'}]=1-\frac{\theta_\Lambda\left(\frac{\xi+\xi'}{2}\right)}{\sqrt{\theta_\Lambda(\xi)\theta_\Lambda(\xi')}}$\\
R\'enyi $\alpha$-divergence & 
$D_\alpha[p_\xi:p_{\xi'}]=\frac{1}{\alpha-1} \log \left(\sum_{l \in \Lambda} p_\xi(l)^{\alpha} p_{\xi'}(l)^{1-\alpha}\right)$
\\ ($\alpha>0, \alpha\not=1$) & $D_\alpha[p_\xi:p_{\xi'}]=\frac{\alpha}{1-\alpha} \log  \frac{\theta_\Lambda(\xi)}{\theta_\Lambda(\alpha\xi+(1-\alpha)\xi')}+
  \log\frac{\theta_\Lambda(\xi')}{\theta_\Lambda(\alpha\xi+(1-\alpha)\xi')}$\\ 
	& $\lim_{\alpha\rightarrow 1} D_\alpha[p_\xi:p_{\xi'}]= D_\KL[p_\xi:p_{\xi'}]$\\
$\gamma$-divergence &
$D_\gamma[p_\xi:p_{\xi'}]=\frac{1}{\gamma(\gamma-1)} 
\log \left(
\frac{\left(\sum_{l\in\Lambda} p_\xi^{\gamma}(x)\right)\, \left(\sum_{l\in\Lambda} p_{\xi'}^{\gamma}(l)\right)^{\gamma-1}}{
\left(\sum_{l\in\Lambda} p_\xi(l) p_{\xi'}^{\gamma-1}(l))
\right)^{\gamma}}\right)$
\\ ($\gamma>1$) &
$D_\gamma[p_\xi:p_{\xi'}]=\frac{1}{\gamma(\gamma-1)} 
\log \left(
\frac{\theta_\Lambda(\gamma\xi)\, \theta_\Lambda(\gamma\xi')^{\gamma-1}}{
\theta_\Lambda(\xi+(\gamma-1)\xi')^{\gamma}
}\right)$
\\ 
& $\lim_{\gamma\rightarrow 1} D_\gamma[p_\xi:p_{\xi'}]=D_\KL[p_\xi:p_{\xi'}]$ \\
H\"older divergence & 
$D_{\alpha, \gamma}^{\mbox{H\"older}}[r: s]:=
\left|\log \left(\frac{\sum_{x\in\mathcal{X}} r(x)^{\gamma / \alpha} s(x)^{\gamma / \beta} }{\left(\sum_{x\in\mathcal{X}} r(x)^{\gamma}\right)^{1 / \alpha}\left(\sum_{x\in\mathcal{X}} s(x)^{\gamma} \right)^{1 / \beta}}\right)\right|$\\
($\gamma>0$, $\frac{1}{\alpha}+\frac{1}{\beta}=1$) & $D_{\alpha, \gamma}^{\mbox{H\"older}}[p_\xi: p_{\xi'}]=\left|\log \frac{\theta_\Lambda(\gamma\xi)^{\frac{1}{\alpha}} \theta_\Lambda(\gamma\xi')^{\frac{1}{\beta}}}
{\theta_\Lambda(\frac{\gamma}{\alpha}\xi+\frac{\gamma}{\beta}\xi')}\right|
$\\
Cauchy-Schwarz divergence & 
$D_{\mathrm{CS}}[r: s]:=
-\log\frac{\sum_{x\in\calX} r(x)s(x)}{\sqrt{(\sum_{x\in\calX} r^2(x))\, (\sum_{x\in\calX} s^2(x)) }}
$\\
(H\"older with $\alpha=\beta=\gamma=2$) & 
$D_{\mathrm{CS}}[p_\xi: p_{\xi'}]=\log\frac{\sqrt{\theta_\Lambda(2\xi)\theta_\Lambda(2\xi')}}{\theta_\Lambda(\xi+\xi')}$\\
\hline
\end{tabular}
}

\caption{Summary of statistical divergences with corresponding formula for lattice Gaussian distributions with partition function $\theta_\Lambda(\xi)$. Ordinary parameterization $\lambda(\xi)=(\mu=E_{p_\xi}[X],\Sigma=\Cov_{p_\xi}[X])$ for $X\sim N_\Lambda(\xi)$.}
\label{tab:summarydiv}
\end{table}
}

Other statistical divergences like the projective H\"older divergences~\cite{nielsen2017holder}  between lattice Gaussian distributions can be obtained  similarly in closed-form:
$$
D_{\alpha, \gamma}^{\mathrm{H}}[r: s]:=
\left|\log \left(\frac{\sum_{x\in\mathcal{X}} r(x)^{\gamma / \alpha} s(x)^{\gamma / \beta} }{\left(\sum_{x\in\mathcal{X}} r(x)^{\gamma}\right)^{1 / \alpha}\left(\sum_{x\in\mathcal{X}} s(x)^{\gamma} \right)^{1 / \beta}}\right)\right|, \left(\gamma>0, \frac{1}{\alpha}+\frac{1}{\beta}=1\right).
$$
The H\"older divergences include the  Cauchy-Schwarz divergence~\cite{jenssen2006cauchy} for $\gamma=\alpha=\beta=2$:
$$
D_{\mathrm{CS}}[r: s]:=
-\log\frac{\sum_{x\in\calX} r(x)s(x)}{\sqrt{(\sum_{x\in\calX} r^2(x))\, (\sum_{x\in\calX} s^2(x)) }}.
$$

Since the natural parameter space $\Xi$ is a cone~\cite{nielsen2017holder}, we get:

$$
D_{\alpha, \gamma}^{\mathrm{H}}[p_\xi: p_{\xi'}]=\left|\log \frac{\theta_\Lambda(\gamma\xi)^{\frac{1}{\alpha}} \theta_\Lambda(\gamma\xi')^{\frac{1}{\beta}}}
{\theta_\Lambda(\frac{\gamma}{\alpha}\xi+\frac{\gamma}{\beta}\xi')}\right|.
$$

Thus we get the following closed-form for the Cauchy-Schwarz divergence between two lattice Gaussian distributions:

$$
D_{\mathrm{CS}}[p_\xi: p_{\xi'}]=
\log\frac{\sqrt{\theta_\Lambda(2\xi)\theta_\Lambda(2\xi')}}{\theta_\Lambda(\xi+\xi')}.
$$

\bibliographystyle{plain}
\bibliography{IGDiscreteGaussianSiegelBIB}

\appendix

%%%
\section{Code snippet in Julia}\label{sec:Julia}
%%%%

The Julia language can be freely downloaded from \url{https://julialang.org/}.

The ellipsoids used to approximate the theta function $\theta_R$ are stored in the {\tt RiemannMatrix} structure of the {\tt Theta.jl} Julia package:

 \begin{center}
\includegraphics[width=0.95\textwidth]{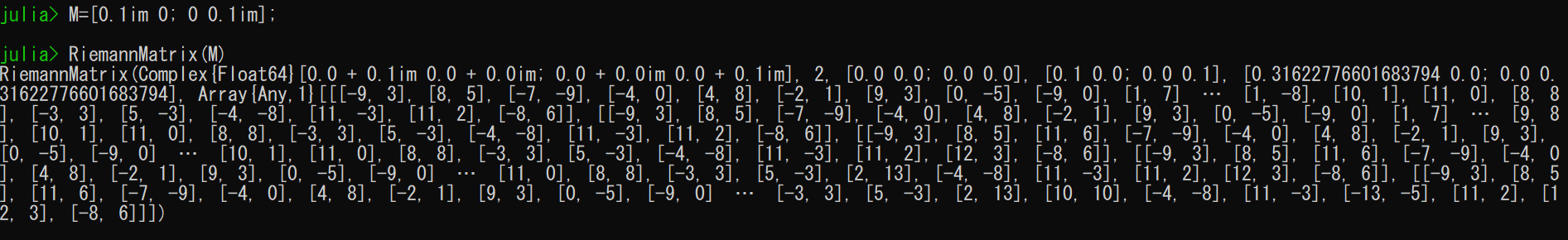}
\end{center}

Executing the code below gives the following result:
\begin{verbatim}
julia> BhattacharyyaDistance(v1,M1,v2,M2)
1.6259948590224578

julia> KLDivergence(v1,M1,v2,M2)
7.841371347366552
\end{verbatim}

\definecolor{ForestGreen}{RGB}{34,139,34}
\begin{lstlisting}
# in Julia 1.4.2
using Theta

M1=[0.1 0; 0 0.2];
v1 = [-0.2;-0.2];
M2=[0.15 0; 0 0.25];
v2 = [0.2;0.2];

# cumulant function of the discrete normal family
function F(v,M)
R= RiemannMatrix(im*M);
log(real(theta(-im*v,R)))
end

# Renyi divergence between two discrete normal distributions
function RenyiDivergence(alpha,v1,M1,v2,M2)
M12=alpha*M1+(1-alpha)*M2;
v12=alpha*v1+(1-alpha)*v2;
(1/(1-alpha))* (alpha*F(v1,M1) +(1-alpha)*F(v2,M2) - F(v12,M12)) 
end

function BhattacharyyaDistance(v1,M1,v2,M2)
(1/2)*RenyiDivergence(1/2,v1,M1,v2,M2)
end

function KLDivergence(v1,M1,v2,M2)
alpha=0.9999999999;
RenyiDivergence(alpha,v1,M1,v2,M2)
end

BhattacharyyaDistance(v1,M1,v2,M2)
KLDivergence(v1,M1,v2,M2)
\end{lstlisting}

\end{document}